\def\~{{$\tilde{\phantom{a}}$}}

\documentclass [12pt] {article}
\usepackage{epsfig}
\usepackage{color}

\def\thebibliography#1{\section{References}\markboth
 {REFERENCES}{REFERENCES}\list
 {[\arabic{enumi}]}{\settowidth\labelwidth{[#1]}\leftmargin\labelwidth
 \advance\leftmargin\labelsep
 \usecounter{enumi}}
 \def\newblock{\hskip .11em plus .33em minus -.07em}
 \sloppy
 \sfcode`\.=1000\relax}
\def\upcite#1{\raise6pt\hbox{\scriptsize
\cite{#1}}}
\def\lsim{\mathrel {\vcenter {\baselineskip 0pt \kern 0pt
    \hbox{$<$} \kern 0pt \hbox{$\sim$} }}}
\def\gsim{\mathrel {\vcenter {\baselineskip 0pt \kern 0pt
    \hbox{$>$} \kern 0pt \hbox{$\sim$} }}}
\def\gtlt{\mathrel {\vcenter {\baselineskip 0pt \kern 0pt
    \hbox{$>$} \kern 0pt \hbox{$<$} }}}

\def\hline{\noalign{\hrule \vskip2pt}}

\def\|{\ifmmode\Vert\else \char`\|\fi}
\ifx\oldzeta\undefined                          
  \let\oldzeta=\zeta                            
  \def\zzeta{{\raise 2pt\hbox{$\oldzeta$}}}     
  \let\zeta=\zzeta                              
\fi

\ifx\oldchi\undefined                           
  \let\oldchi=\chi                              
  \def\cchi{{\raise 2pt\hbox{$\oldchi$}}}       
  \let\chi=\cchi                                
\fi

\def\frac#1#2{{#1 \over #2}}

\def\half{\ifinner {\scriptstyle {1 \over 2}}
   \else {1 \over 2} \fi}

\def\simge{\mathrel{%
   \rlap{\raise 0.511ex \hbox{$>$}}{\lower 0.511ex \hbox{$\sim$}}}}
\def\simle{\mathrel{
   \rlap{\raise 0.511ex \hbox{$<$}}{\lower 0.511ex \hbox{$\sim$}}}}

\def\buildchar#1#2#3{{\null\!                   
   \mathop#1\limits^{#2}_{#3}                   
   \!\null}}                                    
\def\overcirc#1{\buildchar{#1}{\circ}{}}

\def\slashchar#1{\setbox0=\hbox{$#1$}           
   \dimen0=\wd0                                 
   \setbox1=\hbox{/} \dimen1=\wd1               
   \ifdim\dimen0>\dimen1                        
      \rlap{\hbox to \dimen0{\hfil/\hfil}}      
      #1                                        
   \else                                        
      \rlap{\hbox to \dimen1{\hfil$#1$\hfil}}   
      /                                         
   \fi}                                         %

\def\subrightarrow#1{
  \setbox0=\hbox{
    $\displaystyle\mathop{}
    \limits_{#1}$}
  \dimen0=\wd0
  \advance \dimen0 by .5em
  \mathrel{
    \mathop{\hbox to \dimen0{\rightarrowfill}}
       \limits_{#1}}}                           

\def\overlay#1#2{\ifmmode%
\setbox0=\hbox{$#1$}%
\setbox1=\hbox to\wd0{\hss$#2$\hss}\else%
\setbox0=\hbox{#1}%
\setbox1=\hbox to\wd0{\hss#2\hss}\fi%
#1\hskip-\wd0\box1 }

\def\pmb#1{\leavevmode\setbox0=\hbox{#1}%
\kern-.02em\copy0\kern-\wd0
\kern.04em\copy0\kern-\wd0
\kern-.02em\raise.04em\box0 }

\def\vereq#1#2{\lower3pt\vbox{\baselineskip1.5pt \lineskip1.5pt
\ialign{$\m@th#1\hfill##\hfil$\crcr#2\crcr\sim\crcr}}}

\def\tensor#1{\protect\@ontopof{#1}{\leftrightarrow}{1.15}\mathord{\box2}}
\def\overstar#1{\protect\@ontopof{#1}{\ast}{1.15}\mathord{\box2}}
\def\overdots#1{\protect\@ontopof{#1}{\cdots}{1.0}\mathord{\box2}}
\def\overcirc#1{\protect\@ontopof{#1}{\circ}{1.2}\mathord{\box2}}
\def\loarrow#1{\protect\@ontopof{#1}{\leftarrow}{1.15}\mathord{\box2}}
\def\roarrow#1{\protect\@ontopof{#1}{\rightarrow}{1.15}\mathord{\box2}}

\def\@ontopof#1#2#3{%
{\mathchoice
{\@@ontopof{#1}{#2}{#3}\displaystyle\scriptstyle}%
{\@@ontopof{#1}{#2}{#3}\textstyle\scriptstyle}%
{\@@ontopof{#1}{#2}{#3}\scriptstyle\scriptscriptstyle}%
{\@@ontopof{#1}{#2}{#3}\scriptscriptstyle\scriptscriptstyle}%
}%
}

\def\@@ontopof#1#2#3#4#5{%
\setbox0=\hbox{$#4#1$}%
\setbox1=\hbox{$#5#2$}%
\setbox2=\hbox{}\ht2=\ht0 \dp2=\dp0 %
\ifdim\wd0>\wd1 %
\setbox1=\hbox to\wd0{\hss\box1\hss}%
\mathord{\rlap{\raise#3\ht0\box1}\box0}%
\else   %
\setbox1=\hbox to.9\wd1{\hss\box1\hss}%
\setbox0=\hbox to\wd1{\hss$#4\relax#1$\hss}%
\mathord{\rlap{\copy0}\raise#3\ht0\box1}%
\fi
}%

\def\lambdabar{\protect\@lambdabar}
\def\@lambdabar{%
\relax
\bgroup
\def\@tempa{\hbox{\raise.73\ht0
\hbox to0pt{\kern.25\wd0\vrule width.5\wd0
height.1pt depth.1pt\hss}\box0}}%
\mathchoice{\setbox0\hbox{$\displaystyle\lambda$}\@tempa}%
{\setbox0\hbox{$\textstyle\lambda$}\@tempa}%
{\setbox0\hbox{$\scriptstyle\lambda$}\@tempa}%
{\setbox0\hbox{$\scriptscriptstyle\lambda$}\@tempa}%
\egroup
}

\def\corresponds{{\lower.2ex\hbox{=}}{\rm\kern-.75em^\triangle}}
\def\succsim{\succ\kern-.9em_\sim\kern.3em}
\def\precsim{\prec\kern-1em_\sim\kern.3em}
\def\slantfrac#1#2{\kern1em^{#1}\kern-.3em/\kern-.1em_{#2}}

\def\dddot#1{{\buildrel {. \kern-.05em . \kern-.05em .} \over {#1}}}

\begin{document}
                                                                
\begin{center}
{\Large\bf A Capacitor Paradox}
\\

\medskip

Kirk T.~McDonald
\\
{\sl Joseph Henry Laboratories, Princeton University, Princeton, NJ 08544}
\\
(July 10, 2002)
\end{center}

\section{Problem}

Two capacitors of equal capacitance $C$ are connected in parallel by zero-resistance
wires and a switch, as shown in the lefthand figure below.  Initially the switch is open, one
capacitor is charged to voltage $V_0$ and the other is uncharged.  At time $t = 0$
the switch is closed.  If there were no damping mechanism, the circuit would then
oscillate forever, at a frequency dependent on the self inductance $L$ and the capacitance
$C$.  However, even in a circuit with zero Ohmic resistance, damping occurs due to the
radiation of the oscillating charges, and eventually a static charge distribution results.

\vspace{0.1in}
\centerline{\includegraphics[width=5in]{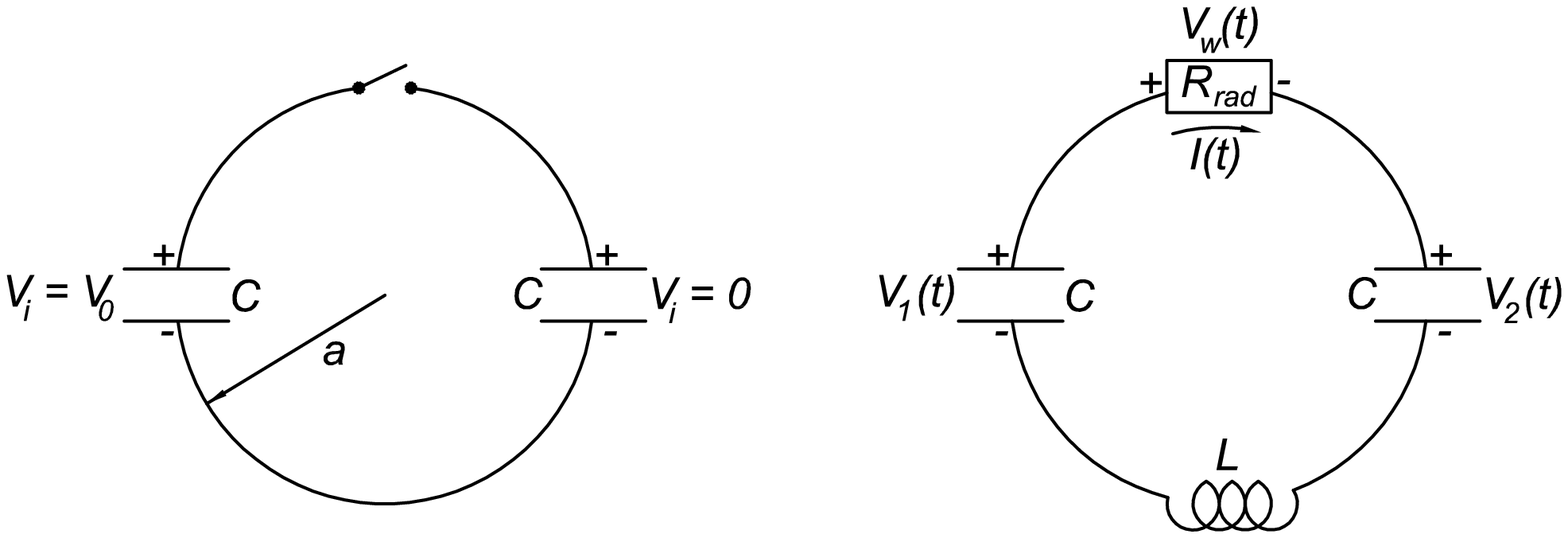}} 

\begin{enumerate}
\item
How much energy is stored in the system before and (a long time) after the switch is closed?
The paradox is that these two energies are not equal.

\item
Verify that the ``missing'' stored energy has been radiated away by the transient 
current after
the switch was closed.  (Recall that in Poynting's view, the energy the energy that
is transferred from one capacitor to the other passes through the intervening space,
not down the connecting wires.)

For this, you may assume that the wires form a circle of radius $a$ and that the conduction
currents in the capacitor plates may be neglected.  In this approximation the current
in the wires is spatially uniform.  You may neglect the charge
accumulation in the wires compared to that on the capacitor plates, in which case the
total electric dipole moment of the system is constant and magnetic dipole radiation
dominates.  Also ignore the self inductance of the circuit.

The ``radiation resistance'' of this
circuit causes a voltage drop $V_w$ to occur along the nominally zero-resistance wires that 
can be identified as
\begin{equation}
V_w(t) = {P_{\rm rad}(t) \over I(t)}\, ,
\label{p1}
\end{equation}
where $P_{\rm rad}$ is the radiated power and $I(t)$ is the current in the wire.  That is,
the radiating wires act like a third circuit element that can be combined with the two
capacitors in an analysis of the voltage and current in the circuit, as sketched in the
equivalent circuit above.

From this circuit analysis deduce a (nonlinear) differential equation for the difference
in voltages on the two capacitors, $V_{12} = V_1 - V_2$, which has an exponential solution.  
Using this solution, verify that the total
radiated energy is equal to the loss of energy stored in the capacitors after the switch
is closed.  Also, give the magnitude of the radiation resistance (in Ohms).

\item
Extend the analysis to include self inductance.

\end{enumerate}

\section{Solution}

This problem is adapted from ``The two-capacitor problem with radiation'' by T.B.~Boylan,
D.~Hite and N.~Singh, Am.\ J.\ Phys.\ {\bf 70}, 415 (2002).

\begin{enumerate}
\item
The energy stored in a capacitor of capacitance $C$ that is charged to voltage $V$ is
$U = C V^2 / 2$.  Hence the initial stored energy of the system is
\begin{equation}
U_i = {C V_0^2 \over 2}\, .
\label{s1}
\end{equation}
A long time after the switch has been closed the initial charge is equally distributed
between the two capacitors, each of which is now at voltage $V_0 / 2$.  Hence, the final
stored energy of the system is
\begin{equation}
U_f = 2 {C (V_0 / 2)^2 \over 2} =  {C V_0^2 \over 4} = {U_i \over 2}\, .
\label{s2}
\end{equation}

\item
To calculate the radiated power we note that the stated approximations are such that 
electric multipole radiation by the circuit may be neglected, and magnetic dipole radiation
dominates.  The magnetic moment $m$ of the circuit is (in Gaussian units)
\begin{equation}
m(t) = {\pi a^2 I(t) \over c}\, ,
\label{s3}
\end{equation}
where $c$ is the speed of light.  According to the Larmor formula, the radiated power is
\begin{equation}
P_{\rm rad} = {2 \ddot m^2 \over 3 c^3} = {2\pi^2 a^4 \ddot I^2 \over 3 c^5}\, .
\label{s4}
\end{equation}
As suggested, we introduce the voltage drop $V_w$ along the wires due to the
radiation resistance via
\begin{equation}
V_w = {P_{\rm rad} \over I} = {2\pi^2 a^4 \ddot I^2 \over 3 c^5 I}\, .
\label{s5}
\end{equation}
The current $I$ is also related to the charge and voltage on the capacitors by
\begin{equation}
\dot V_1 = {\dot Q_1 \over C} = - {I \over C}\, , \qquad \mbox{and} \qquad
\dot V_2 = {\dot Q_2 \over C} =  {I \over C}\,
\label{s7}
\end{equation}
An additional relational between the currents and voltages is obtained from
Kirchhoff's circuit law,
\begin{equation}
V_1 - V_2 + L \dot I - V_w = 0,
\label{s6}
\end{equation}
including the self-inductance term for later consideration.

Our strategy now is to use $V_{12} = V_1 - V_2$ as the independent variable,
whose initial value is $V_0$.  From eq.~(\ref{s7}) we have
\begin{equation}
I = - {C \dot V_{12} \over 2}\, , \qquad \mbox{and\ hence}, \qquad
\dot I = - {C \ddot V_{12} \over 2}\, , \qquad
\ddot I = - {C \dddot V_{12} \over 2}\, .
\label{s8}
\end{equation}
Using eqs.~(\ref{s5}) and (\ref{s8}) in eq.~(\ref{s6}) we find the 
differential equation for $V_{12}$ to be
\begin{equation}
\dddot V_{12}^2 + {3 c^5 L\over 2 \pi a^4} \dot V_{12} \ddot V_{12}   
 + {3 c^5 \over \pi a^4 C} V_{12} \dot V_{12} = 0. 
\label{s9}
\end{equation}
We try a solution of the form
\begin{equation}
V_{12}(t > 0) = V_0 e^{-t / \tau},
\label{s10}
\end{equation}
which satisfies eq.~(\ref{s9}) provided
\begin{equation}
{1 \over \tau^6} = {3 c^5 L \over 2 \pi a^4 \tau^3} +{3 c^5 \over \pi a^4 C \tau}\, .
\label{s11}
\end{equation}

We first suppose that we may set the self inductance $L$ to zero.
Of the six possible solutions for $\tau$ in this case, we choose the only one that is real and
finite,
\begin{equation}
\tau = \left({\pi a^4 C \over 3 c^5}\right)^{1/5}.
\label{s12}
\end{equation}
The capacitance $C$ in eq.~(\ref{s12}) is representative of the total capacitance of the
system, which can never be much smaller than length $a$, since the capactiance of the
wires alone has roughly this value.  Hence, the time constant of the discharge
of the capacitors obeys $\tau \geq a / c$.  That is, the discharge time is longer than
the transit time of light across the circuit.

The total radiated power can now be calculated, combining eqs.~(\ref{s4}), (\ref{s8}),
(\ref{s10}) and (\ref{s12}),
\begin{equation}
U_{\rm rad} = \int_0^\infty P_{\rm rad}\ dt
= \int_0^\infty {2\pi^2 a^4  \over 3 c^5} 
{C^2 V_0^2 e^{-2 t / \tau} \over 4  \tau^6} dt
= {\pi^2 a^4  \over 3 c^5} {C^2 V_0^2  \over 4  \tau^5}
= {C V_0^2  \over 4}
= U_i - U_f,
\label{s13}
\end{equation}
and is indeed equal to the ``missing'' stored energy.

Finally, we identify the radiation resistance by combining eqs.~(\ref{s6}), (\ref{s8}) 
and (\ref{s10}) (ignoring the self inductance) in the form
\begin{equation}
V_w = {2 \tau \over C} I \equiv I R_{\rm rad}.
\label{s14}
\end{equation}
Thus,
\begin{equation}
R_{\rm rad} = {2 \tau \over C} 
= {1 \over c} \left({\pi \over 3}\right)^{1/5} \left({a \over C}\right)^{4/5} 
\lsim {1 \over c} = 30 \Omega.
\label{s15}
\end{equation}

Since this value is large compared to the resistance of typical lead wires in a circuit,
our approximation of zero-resistance wires is a good one (unless $C \gg a$).

\item
In extending the analysis to include self induction, it may be useful to first recall the
behavior of the circuit if radiation is neglected.  Then, the circuit equation (\ref{s9})
reduces to the form
\begin{equation}
\ddot V_{12} + {2 \over L C} V_{12} = 0, 
\label{s101}
\end{equation}
which has the oscillatory solution (for the stated initial conditions)
\begin{eqnarray}
Q_1 & = & {C V_0 \over 2} (1 + \cos \omega t),
\label{s102} \\
Q_2 & = & {C V_0 \over 2} (1 - \cos \omega t),
\label{s103} \\
I & = & - \dot Q_1 = \dot Q_2 =  {C V_0 \omega \over 2} \sin \omega t,
\label{s104} \\
\end{eqnarray}
where the oscillation frequency is
\begin{equation}
\omega = \sqrt{2 \over LC}\, .
\label{s105}
\end{equation}
The stored energy is
\begin{equation}
U = {1 \over 2} L I^2 + {Q_1^2 + Q_2^2 \over 2 C} = {C V_0^2 \over 2} = U_0,
\label{s106}
\end{equation}
which is constant in time at its initial value, since there is no dissipative mechanism
by assumption.

The self inductance of a circular loop (a torus) of wire of thickness $2b$ (minor radius 
$b$ and major radius $a \gg b$ is given by
 \begin{equation}
L = {4 \pi a \over c^2} \left( \ln {8 a \over b} - {7 \over 4} \right).
\label{s16}
\end{equation}
For $a / b \approx 100$, the self inductance is $L \approx 60 a / c^2$.  The LC oscillation
of the circuit has time constant $\approx \sqrt{LC} \approx \sqrt{60 a C / c^2}
\approx 8 a / c \approx 8 \tau_{\rm rad}$, where we again suppose that $C \approx a$.
Thus, for reasonable circuit parameters, the radiation damping time is of the same order
of magnitude as the $LC$ oscillation period.  In practice, we can have either
over- or underdamped oscillations, depending on the parameter values. 

In the case of overdamped ``oscillations'', the self inductance is largely ignorable and the
analysis given in item 2 is still valid.  For weakly damped oscillations, the second term
on the righthand side of eq.~(\ref{s11}) is small compared to the first, so an iterative
solution for $1/\tau$ follows readily.  But as the details are not particularly illuminating
we leave them to readers more motivated by practice than by principle.

\end{enumerate}

\end{document}